\documentclass[iicol,sn-mathphys-num]{sn-jnl}
\usepackage{amsmath,amsfonts}
\usepackage{hyperref}
\usepackage[title]{appendix}
\hypersetup{colorlinks=true, urlcolor=blue,linkcolor=blue, citecolor=blue}
\usepackage[utf8]{inputenc}
\usepackage{caption}
\usepackage{subcaption}
\usepackage{graphicx}
\usepackage{braket}
\usepackage{natbib}

\begin{document}    
	
	\title{Instantons and transseries of the Mathieu potential deformed by a $\mathcal{PT}$ symmetry parameter}
	
	\author*[1,2]{\fnm{Natália} M.\@ \sur{Alvarenga}}
        \email{natalia\_de\_melo@msn.com}

	\author[3]{\fnm{Erich} \sur{Cavalcanti}}
	\email{ecav@pm.me}
	 
	\author[2]{\fnm{Cesar}  A.\@ \sur{Linhares}}
	\email{linharescesar@gmail.com}

	\author[4]{\fnm{José André} \sur{Lourenço}}
	\email{jose.lourenco@ufes.br}
	 
	\author[2]{\fnm{José Roberto} P.\@ \sur{Mahon}}
	\email{mahon@uerj.br}

        \author[5]{\fnm{Fernando} P.\@ P.\@ \sur{Reis}}
        \email{fernando.reis@ufes.br}

\affil*[1]{\orgdiv{Instituto de Física}, \orgname{Universidade de Brasília}, \postcode{70910-900}, \city{Brasília}, \state{DF}, \country{Brazil}; ORCID 0000-0002-2637-1323}
\affil[2]{\orgdiv{Instituto de Física}, \orgname{Universidade do Estado do Rio de Janeiro}, \postcode{20550-013}, \city{Rio de Janeiro}, \state{RJ}, \country{Brazil}}
\affil[3]{\orgname{Centro Brasileiro de Pesquisas F\'{\i}sicas/MCTI}, \postcode{22290-180}, \city{Rio de Janeiro}, \state{RJ}, \country{Brazil}}
\affil[4]{\orgdiv{Departamento de Ciências Naturais}, \orgname{Universidade Federal do Espírito Santo}, \postcode{29932-540}, \city{São Mateus}, \state{ES}, \country{Brazil}}
\affil[5]{\orgdiv{Departamento de Matemática Aplicada}, \orgname{Universidade Federal do Espírito Santo}, \postcode{29932-540}, \city{São Mateus}, \state{ES}, \country{Brazil}}

\abstract{
We investigate the non-perturbative effects of a deformation of the Mathieu differential equation consistent with $\mathcal{PT}$ symmetry. First, we develop a connection between the non-Hermitian and Hermitian scenarios by a reparameterization in the complex plane,
followed by a restriction of the $\mathcal{PT}$ deformation parameter. The latter is responsible for preserving the information about $\mathcal{PT}$ symmetry when we choose to work in the Hermitian scenario. We note that this factor is present in all non-perturbative results and in the transseries representation of the deformed Mathieu partition function that we have obtained. In quantum mechanics, we found that the deformation parameter of $\mathcal{PT}$ symmetry has an effect on the real instanton solution for the deformed Mathieu potential in the Hermitian scenario. As its value increases, the non-Hermiticity factor makes it smoother for the instanton to pass from one minimum to another, that is, it modifies the instanton width. The explanation for this lies in the fact that the height of the potential barrier decreases as we increase the value of the deformation parameter. We present how this effect extends to the multi-instanton level and to the bounce limit of an instanton-anti-instanton pair. As an application of the obtained results, we show that the equation of motion under a tilted version of the potential in the Hermitian scenario compares to the resistively shunted junction (RSJ) model for the Josephson junction.} 



	\maketitle

\section{Introduction} \label{sec:introduction}
The Mathieu equation is a special case of a linear second-order homogeneous differential equation, which occurs in many applications in physics and engineering. Historically, it is related to the investigations of the French mathematician É. L. Mathieu on the free oscillations of an elliptical membrane~\cite{bib:Mathieu68}. Over time, a mathematical theory for the Mathieu equation and the related Mathieu functions has been developed and applied in several contexts~\cite{bib:Mclachlan,bib:aplications,NIST:DLMF}. Some physical examples are: electromagnetic and elastic wave equations with elliptical boundary conditions (as in waveguides or resonators)~\cite{bib:elliptical}; motion of particles in alternated-gradient focusing~\cite{bib:alternating} or electromagnetic traps~\cite{bib:trap}; inverted pendulum; parametric oscillators; motion of a quantum particle in a periodic potential~\cite{bib:Carver}; and eigenfunctions of quantum pendulum~\cite{bib:QPendulum}. In particular, the Mathieu equation with a complex driven parameter was used to describe  charge carriers moving in a sinusoidally curved graphene~\cite{bib:Kerner}. Interesting resurgent features of the Mathieu potential have been explored in the literature~\cite{Cherman:2014ofa,Cherman:2014xia,Basar:2013eka} from the analysis of divergent asymptotic series that arise when expanding the Mathieu partition function, both perturbatively and non-perturbatively.\footnote{The potential in Ref.~\cite{Basar:2013eka} is a generalization that interpolates between trigonometric (Mathieu) and hyperbolic forms.}

Locating a point in space-time involves using a four-dimensional vector $(x,y,z,t)$. The parity operator $\mathcal{P}$ affects the sign of the spatial part of this four-vector, while the time-reversal operator $\mathcal{T}$ changes only the sign of its time component. Together, the total effect of parity and time ($\mathcal{PT}$) reflection is to change the sign of all its components, namely, $(x,y,z,t)\rightarrow (-x,-y,-z,-t)$. Furthermore, there is another interesting effect when it comes to quantum mechanics, due to the fact that both the position and momentum operators obey the Heisenberg algebra ($xp-px=i$). Although $\mathcal{P}$ keeps the Heisenberg algebra unchanged, the application of $\mathcal{T}$ leads to a change in the  momentum $p$ sign, forcing the imaginary unit $i$ to change its sign to preserve the Heisenberg algebra under $\mathcal{PT}$.\footnote{Wigner has shown~\cite{Wigner:1932operation} that, in quantum mechanics, the time-reversal operator $\mathcal{T}$ changes the sign of the imaginary number $i$.} From this, we can assert that a Hamiltonian is $\mathcal{PT}$-symmetric when the condition $V(\mathbf{x})=V^{*}(-\mathbf{x})$ is satisfied~\cite{bib:Optical}.

In textbook quantum mechanics, one requires the Hamiltonian operator to be Hermitian in order to guarantee real eigenvalues. However, space-time $\mathcal{PT}$-symmetric systems can describe physical models~\cite{bib:Bender1998}. There are many experimental applications~\cite{Bender:2016epn}.  
Furthermore, $\mathcal{PT}$-symmetric Hamiltonians are suitable for investigating the boundary between open and closed systems~\cite{bib:livroptbender}. This is indicated by looking at the spectrum, in which we can have a $\mathcal{PT}$-symmetric phase (where all energy levels are real) and a $\mathcal{PT}$-broken phase (where some of the energy levels become complex), which is controlled by some strain parameter in the Hamiltonian. Such change occurs because the $\mathcal{PT}$ operator is non-linear. Therefore, although it commutes with the Hamiltonian, there is no guarantee that they will share the same eigenstates. This richness has motivated a multitude of physical applications in optics, superconductivity, microwave cavity, lasers, electronic circuits, chaos, graphene, metamaterials etc., and the research on the subject has experienced an overwhelming growth in recent years~\cite{Bender:2016epn}. In Appendix \ref{ap:A}, we discuss the role of a $\mathcal{PT}$ symmetry operator and how it can be understood in terms of an underlying inner-product space, as well as the conditions for the reality of the spectrum.

In the present paper, we explore some non-perturbative effects of a deformation of the Mathieu differential equation consistent with $\mathcal{PT}$ symmetry. The aspects related to its spectrum and the transition from $\mathcal{PT}$-unbroken to $
\mathcal{PT}$-broken phases are numerically
studied by the authors in Ref.~\cite{Cavalcanti:2022wnh}. 

The path to solve the Mathieu differential equation is to employ perturbation theory and obtain a perturbative series, but perturbative series are usually divergent in physics~\cite{bib:zinnjustin,bib:livromarino,bib:artigomarino}. When the series is asymptotic, it is common to employ Borel summation to extract non-perturbative information. However, a more difficult task is the reconstruction of functions from divergent perturbative expansions, as it generates incomplete information. This signals the need to include non-perturbative instanton effects. Such divergence is related to the existence of singularities in the Borel complex plane, usually also associated with instantons~\cite{Cherman:2014ofa,bib:artigomarino}. These are crucial parts in the construction of transseries, which are important manipulations of formal series that can cancel out ambiguities~\cite{Cherman:2014ofa,Cherman:2014xia,bib:artigomarino,Basar:2013eka,ANICETO20191,ANICETO2015}.

Instantons were discovered in 1975 in the context of quantum chromodynamics~\cite{bib:Polyakov1975,bib:ABCinstantons}, being understood as topologically non-trivial classical solutions of field equations, that is, they relate different vacua/minima~\cite{bib:livromarino,bib:Das,bib:coleman,Zinn-Justin:1982aya,bib:ZJCritical}. Usually, the subject is presented in the context of an analytic continuation to imaginary time (Wick rotation) and conceptualize instantons as solutions of classical equations of motion in Euclidean spacetime that have finite (non-zero) action. They are also related to a semi-classical description of quantum tunneling. Finally, another feature that interests us is the classification of instantons as a type of non-perturbative effect, that is, one that cannot be seen in perturbation theory in quantum field theory or quantum mechanics~\cite{bib:livromarino}. 

We start by describing the usual Hermitian case, in Sec.\@ \ref{sec:usual}, then move on to the connection between Hermitian and non-Hermitian scenarios in Sec.\@ \ref{sec:connection}. In Sec.\@ \ref{sec:resurgent} we obtain the resurgent Mathieu partition fuction, and we construct the instanton solution of the deformed Mathieu equation in Sec.\@ \ref{sec:instdef}. In Sec.\@ \ref{sec:dilute} and Sec.\@ \ref{sec:washboard}, we show applications of this formalism to the problems of a dilute gas of deformed instantons.

\section{Usual Hermitian case} \label{sec:usual}
We start with the well-known Mathieu equation~\cite{NIST:DLMF}
\begin{equation} \label{eq:Mathieu}
\left\{-\frac{d^2}{dz^2} +2q\cos(2z)\right\}\psi(z) = a \psi(z),
\end{equation}
with real coefficients $q$ (Mathieu parameter) and $a$ (characteristic number) and whose potential can be written as
\begin{equation} \label{eq:deltazero}
V_U(z)=\sin^2(z)
\end{equation}
by trivial trigonometric identities, for convenience. The following ordinary integral 
\begin{equation} \label{eq:1}
Z_U(g^2)=\frac{1}{g\sqrt{\pi}}\int_{-\pi/2}^{\pi/2} dz \; \exp\left( -\frac{1}{g^2} \sin^2(z) \right),
\end{equation}
where $g$ is a real coupling constant, formally can be considered as a zero-dimensional prototype of the semi-classical partition function approach in path integrals. The purpose of this section is to illustrate the methodology that one can use in order to write the related transseries. This characterizes a zero-dimensional prototype for resurgence~\cite{Tesenat,Basar:2013eka,Cherman:2014ofa,Cherman:2014xia}. Briefly, the saddle points and their associated actions are calculated. The former are easily obtained by taking the derivative of the potential and bringing the result to zero. Two values are found, $z_p=0$ and $z_{np}=\pi/2$, which we call respectively perturbative and non-perturbative saddles. In this zero-dimensional prototype, we simply substitute each saddle point in the expression of the potential (\ref{eq:deltazero}) in order to calculate their associated actions, resulting in $S_p=0$ and $S_{np}=1$. Then comes the main task, which is the perturbative expansion around each saddle of the partition function (\ref{eq:1}). These elements can be inserted into the general structure of a transseries~\cite{Basar:2013eka,Cherman:2014xia}. Some papers on resurgence have already worked with this partition function, which therefore have solid evidence for its resurgent structure~\footnote{Ref.~\cite{Basar:2013eka} uses a potential in terms of $\textrm{sd}(g\phi | m)$ which is a Jacobi elliptic function. Making $m=0$ this falls into the Mathieu potential.}\cite{Cherman:2014ofa,Cherman:2014xia,Basar:2013eka,bib:Misumi}. In the present case, we notice that there are two saddles points, so we can express the transseries as follows,
\begin{equation} \label{eq:transserieusual}
Z_U(g^2)=\sigma_p \, e^{-S_p/g^2} \Phi_p(g^2)+\sigma_{np} \, e^{-S_{np}/g^2} \Phi_{np}(g^2),
\end{equation}
where $\sigma_{p,np}$ are Stokes constants and $S_{p,np}$ are the actions on different perturbative (p) and non-perturbative (np) saddles. The formal series obtained through the perturbative expansion around each saddle are represented by $\Phi_{p,np}$. These can be calculated from the partition function in different ways. It is important to clarify that Stokes constants are not obtained directly from the integral in (\ref{eq:1}). The method used to calculate them is well known in the literature on resurgence~\cite{Cherman:2014ofa,Cherman:2014xia,bib:artigomarino,Tesenat}. When the coefficients of the asymptotic series in (\ref{eq:transserieusual}) are $a_{p,np}(k) \sim k!$, we calculate the  Borel transforms, 
\begin{equation}
B\Phi_{p,np}(t)=\sum_{k=0}^\infty \frac{a_{p,np}(k)}{k!} t^n.
\end{equation}
In the following, we make the $S\Phi_{p,np}(g)$ Borel sum (when it exists) of the formal power series, which is given by the integral of the analytical continuation of $B\Phi_{p,np}(t)$, denoted by $\widetilde{B\Phi_{p,np}}(t)$, see ~\cite{Cherman:2014ofa,Cherman:2014xia,bib:artigomarino,Tesenat}. In the standard Borel summation\footnote{Some authors choose to call the Borel summation the Borel resummation.}, the integration is performed from $t = 0$ to $t=+\infty$ along the real axis, 
\begin{equation}
S\Phi_{p,np}(g) = \frac{1}{g} \int_0^{+\infty} dt \; e^{-\frac{t}{g}} \widetilde{B\Phi_{p,np}}(t),
\end{equation}
When we do not find singularities in the semi-axis $\mathbb{R}^{+}$, this $S\Phi_{p,np}(g)$ exists and it is a function which has the same asymptotic expansion as $\Phi_{p,np}$, but it is well-defined in some finite neighborhood of $g = 0$. However, when we find some singularity in the path or even branch-cut singularities, we say that we have a non-Borel summable series. The lines $\arg(g):=\theta$ where branch-cut singularities are located are called Stokes rays. It is possible to define a generalized Borel sum (also called a lateral or directional Borel (re)summation) that avoids such Stokes rays~\cite{bib:artigomarino}, 
\begin{equation}
S_\theta \Phi_{p,np}(g) =\frac{1}{g}\int_0^{+\infty e^{i\theta}} dt \, e^{-\frac{t}{g}} \widetilde{B\Phi_{p,np}}(t).
\end{equation}
It is also possible to deform the contour so as to enclose the singularities (Hankel contour)~\cite{Cherman:2014ofa}. Stokes constants are factors that arise from the result of the Borel sum, whether standard or generalized\cite{bib:artigomarino}.

In recent work~\cite{Tesenat}, both Mathieu (sine-Gordon) and modified Mathieu (sinh-Gordon) potentials are explored, showing that it is possible to perform computations from the perturbative saddle by writing the partition function in terms of Bessel functions, whose expansions have already been widely explored in the literature. We also offered as alternatives the Newton's generalized binomial theorem followed by Watson's lemma or the integral definition of gamma functions and the direct expansion of the potential in Taylor series. Otherwise, the perturbative expansion around the non-perturbative saddle involves Lefschetz thimbles~\cite{Cherman:2014ofa,Basar:2013eka,Tanizaki,Italianos}, which are a generalization of the steepest-descent method~\cite{bib:livrobender}.

After such procedures, the transseries assumes the form
\begin{eqnarray} \label{eq:transzero}
Z_U(g^2)= \sigma_0 \,  \sum_{k=0}^{\infty} & & a_{0}(k)  g^{2k} + \sigma_1 \, e^{-1/g^2} \nonumber \\
& & \times \sum_{k=0}^{\infty}(-1)^{k}a_{0}(k) g^{2k}, 
\end{eqnarray}
where 
\begin{equation} \label{eq:azero}
a_0=\frac{\left[\Gamma(k+\frac{1}{2})\right]^2}{\Gamma(k+1)\left[\Gamma(\frac{1}{2})\right]^2}
\end{equation}
and $\sigma_0=1$, $\sigma_1=-i$ for $\theta\in \, ]0,\pi[$; also $\sigma_0=1$, $\sigma_1=+i$ for $\theta\in \, ]-\pi,0[$. See Refs.~\cite{Cherman:2014ofa,Cherman:2014xia,Tesenat}.

In one-dimensional quantum mechanics, the partition function is a path integral~\cite{Basar:2013eka,bib:coleman},
\begin{equation} 
\mathcal{Z}_U=\int \mathcal{D}x_U  \exp\left\{ -\int_{-\frac{\beta}{2}}^{\frac{\beta}{2}} dt \left( \frac{m\dot{x}_U^2}{2} +\frac{V_U}{g^2} \right)\right\},
\end{equation}
where $\mathcal{D}x_U$ means integration over all $x_U(t)$ that obey the boundary conditions $x_U(-\beta /2)=x_{Ui}$, $x_U(\beta /2)=x_{Uf}$. The associated instanton equation~\cite{bib:coleman,bib:livromarino,Tesenat} is
\begin{equation} \label{eq:definstanton}
 \dot{x}_U(t)=\sqrt{\frac{2}{m g^2} V_U(x_U(t))}.
\end{equation} 
When we fix $m=1$, $g=1$, it has a real solution 
\begin{equation}
x_U=2\arctan(\exp(\sqrt{2}t)),
\end{equation}
which interpolates between $0$, $\pi$ (see the dotted curve in Fig.~\ref{fig:instanton3deltas}). The imaginary part of the complete solution was neglected as it is constant. The instanton action~\cite{bib:coleman,Tesenat} is defined as 
\begin{equation}
    S_{U}(x_{U})=\int_{-\beta /2}^{\beta/2} dt \, \left(\frac{m}{2} (\dot{x}_{U}(t))^2+\frac{V_U(x_U(t))}{g^2}\right),
    \end{equation}
but it can be written in another way, using (\ref{eq:definstanton}), so that
\begin{eqnarray}
S_{U}(x_{U})&=& \int_{x_{Ui}}^{x_{Uf}} dx_U \, m \dot{x}_U \nonumber \\ 
S_{U}(x_{U}) &=& m \int_{x_{Ui}}^{x_{Uf}} dx_U \, \sqrt{\frac{2V_U(x_U)}{m g^2}}.  
\end{eqnarray} 
In this specific case, keeping $m=1$, $g=1$, knowing that $x_{Ui}=0$, $x_{Uf}=\pi$ and $V_U$ is given by Eq.~(\ref{eq:deltazero}), we get
\begin{equation}
S_U=\sqrt{2}\int_0^\pi \sin(x_U)dx_U=2\sqrt{2}.
\end{equation} 
Real instanton solutions are associated with real saddles and actions whose values are positive. If the reader is interested in knowing about complex saddles and their complex or negative actions, unlike our case, see~\cite{Basar:2013eka}.
 
In the next sections, we will extend the formalism to the case of Mathieu deformed by a non-Hermiticity factor.

\section{Connecting Hermitian and non-Hermitian scenarios}
	\label{sec:connection}
Let us now consider a $\mathcal{PT}$-symmetric version of Eq.(\ref{eq:Mathieu}) in complex domain,  
\begin{equation} \label{eq:deformed}
	\left\{-\frac{d^2}{dz^2} +2q \left[\cos(2z)+i\delta \sin(2z)\right]\right\}\psi(z) = a \psi(z),
\end{equation}
where the parameters $q,\delta,a \in \mathbb{R}$. The case $\delta=0$ leads to the original Mathieu equation (\ref{eq:Mathieu}). The potential $V(z,\delta)=\cos(2z)+i\delta \sin(2z)$ is non-Hermitian and it preserves $\mathcal{PT}$ symmetry for each $\delta$.\footnote{Notice that $\textrm{Re}(V(z,\delta))$ is an \emph{even} function of position, whereas the imaginary part of the potential $V(z,\delta)$ is an \emph{odd} function of position, satisfying the condition $V(z,\delta)=V^{*}(-z,\delta)$.} Equation \eqref{eq:deformed} is known as a generalized $\mathcal{PT}$-symmetric Mathieu equation. In quantum mechanics, we cannot find an analytical solution to the instanton equation in the non-Hermitian scenario, but such solution in the usual Hermitian case is known in the literature. Hence arises the importance of finding a way to rewrite the Mathieu $\mathcal{PT}$-symmetric potential in terms of cosine or in some equivalent form.

\begin{figure}[h!]
	\centering
	\includegraphics[width=0.95\linewidth]{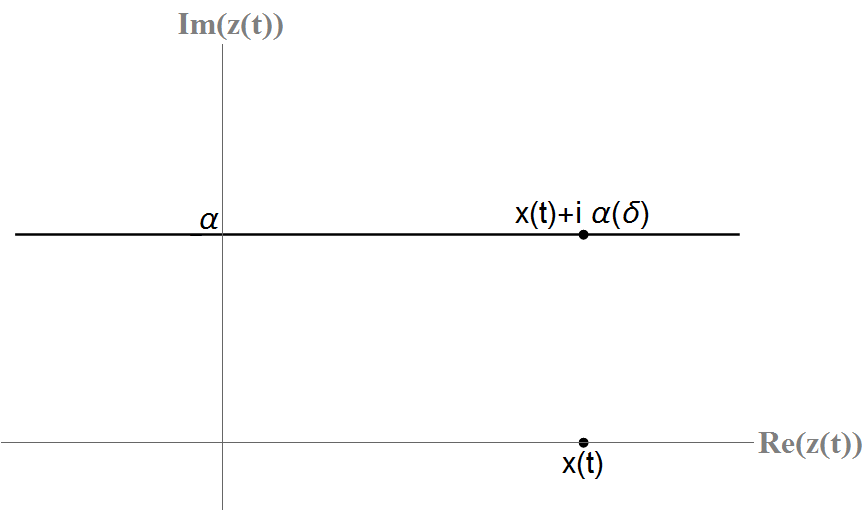}
	\caption{In the restriction $z=x+i\alpha(\delta)$, the imaginary part of $z(t)$ is fixed. Wolfram Mathematica 11.3 was used to create the artwork}
	\label{fig:reparametrizacao}
\end{figure}

If we consider the restriction $z=x+i\alpha(\delta)$ for \eqref{eq:deformed}, which is nothing more than a reparametrization of a line in the complex plane (Fig.\@ \ref{fig:reparametrizacao}), the potential becomes 
\begin{eqnarray}
V(x+i\alpha, & &\delta)= \cos(2x)\, [\text{cosh}(2\alpha)-\delta\text{sinh}(2\alpha)] \nonumber\\ 
& &+i\sin(2x)\, [-\text{sinh}(2\alpha)+\delta \, \text{cosh}(2\alpha)], \nonumber\\ & & \label{eq:vxialpha}
\end{eqnarray} 
where we have used basic trigonometric identities. We can take the imaginary part of (\ref{eq:vxialpha}) as null, allowing us to determine $\alpha:\; \left]-1,1\right[ \; \to \mathbb{R}$, 
\begin{equation}\label{alpha}
\alpha(\delta)=\frac{1}{2}\text{arccosh}\left(\sqrt{\frac{1}{1-\delta^2}}\right).
\end{equation}  
Therefore, the map $\mathcal{V}:\mathbb{R}\times \; \left]-1,1\right[ \; \to \mathbb{R}$, given by
\begin{equation} \label{eq:potencialz}
	\mathcal{V}(x,\delta)=\sqrt{1-\delta^2}\, \sin^2(x),
\end{equation}
for each $\delta\in \; \left]-1,1\right[ \;$, gives us a well-defined Hermitian potential. Furthermore, for each fixed $|\delta|<1$, we can rewrite \eqref{eq:deformed} as
\begin{equation} \label{eq:Hermitian}
\left[\frac{d^2}{dx^2} +4Q\sin^2(x)\right]\psi(x+i\alpha)=(2Q-a) \psi(x+i\alpha),
\end{equation}
where $Q\equiv q\sqrt{1-\delta^2}$ and $\alpha$ is given by \eqref{alpha}. These steps allow us to connect both Hermitian and non-Hermitian scenarios. In Appendix \ref{ap:A}, it is shown that there is an alternative way to rewrite the generalized $\mathcal{PT}$-symmetric Mathieu potential in a cosine form. It shows that there is no obligation to switch to an axis parallel to the real line. The key point of both methods is the restriction on the values of the $\delta$ deformation parameter. We have chosen to use the first method for convenience.

One may think about another perspective, in which the choice is not to nullify the imaginary part of the potential $V(x+i\alpha,\delta)$ in (\ref{eq:vxialpha}), but its real part. In this situation, the equation that establishes the relation between $\alpha$ and $\delta$ parameters is $2\alpha(\delta)=\arctan(i/\delta)$. This corresponds to the complex deformed potential $\mathcal{V}_{Im}(x,\delta)=i\sin(2x)\sqrt{\delta^2-1}$ with $|\delta|>1$. It is possible to move the complex number $i$ into the square root, so that it becomes comparable to that with $|\delta|<1$. Briefly,
\begin{subequations}
\begin{align}
\mathcal{V}(x,\delta)&=\cos(2x)\sqrt{1-\delta^2}, \quad |\delta|<1, \\
\mathcal{V}_{Im}(x,\delta)&=\sin(2x)\sqrt{1-\delta^2}, \quad |\delta|>1.
\end{align}
\end{subequations}
As $\sin(2x+\pi/2)=\cos(2x)$, it is possible to summarize both constraints for delta in a single condition, $|\delta| \neq 1$.

\section{Resurgent Mathieu partition function with $\delta$-deformation} \label{sec:resurgent}
Since the Hermitian scenario of the deformed Mathieu case just has some constant factor multiplied by the function $\sin^2(x)$, whose resurgent structure is well known, it follows immediately that it can be written as a transseries. From \eqref{eq:potencialz} we can compute the saddles depending on $\delta$. That is, $S_p=\mathcal{V}(0,\delta) =0$ and $S_{np}=\mathcal{V}(\frac{\pi}{2},\delta) =\sqrt{1-\delta^2}$. Therefore, the zero-dimensional deformed prototype for the partition function is
\begin{eqnarray}
	Z(\delta, g^2)&=&\frac{2(1-\delta^2)^{1/4}}{g\sqrt{\pi}} \nonumber \\ 
 & & \times \int_{0}^{\pi/2} dx \, \exp\left(\frac{-\sqrt{1-\delta^2}}{g^2}\sin^2(x)\right). \nonumber \\ & &
\end{eqnarray}
Making the change $\sin^2(x)=s$ and writing
\begin{equation}
    (1-s)^{-1/2}=\sum_{k=0}^{\infty}\frac{\Gamma\left(k+\frac{1}{2}\right)}{\Gamma(k+1)\Gamma\left(\frac{1}{2}\right)} s^k,
\end{equation}
we have
\begin{eqnarray}
	Z(\delta, g^2)&=&\frac{(1-\delta^2)^{\frac{1}{4}}}{g\Gamma(\frac{1}{2})}  \sum_{k=0}^{\infty}\frac{\Gamma(k+\frac{1}{2})}{\Gamma(k+1)\Gamma(\frac{1}{2})} \nonumber \\ & & \times \int_{0}^{1} ds \, s^{k-\frac{1}{2}}\exp\left({s\frac{\sqrt{1-\delta^2}}{g^2}}\right).
\end{eqnarray} 
Watson's lemma is then applied to obtain  
\begin{equation}
	Z(\delta, g^2)  \sim \sum_{k=0}^{\infty}\frac{\left[\Gamma(k+\frac{1}{2})\right]^2}{\Gamma(k+1)\Gamma(\frac{1}{2})^2}\left(\frac{\sqrt{1-\delta^2}}{g^2}\right)^{-k}.
\end{equation} 
Finally, we have the transseries representation for the $\delta$-deformed Mathieu partition function,
\begin{align}
	Z(\delta, g^2)= & \sigma_0 \, e^{-S_{p}/g^2}\sum_{k=0}^{\infty}a_{0}(k) A^{-k}(\delta,g^2) \nonumber\\
	&+ \sigma_1 \, e^{-A(\delta,g^2)}\sum_{k=0}^{\infty}(-1)^{k}a_{0}(k) A^{-k}(\delta,g^2),
\end{align}
where
\begin{equation} \label{eq:amp}
A(\delta,g^2)=\frac{1}{g^2}\sqrt{1-\delta^2}.
\end{equation}
Furthermore, we observe that $\sigma_0$ and $\sigma_1$ Stokes constants are the same as those appearing in the transseries representation for the usual Hermitian case (\ref{eq:transzero}). As expected, this transseries becomes identical to (\ref{eq:transzero}) when $\delta$ vanishes. From (\ref{eq:amp}), we can assert that the $g^2$ coupling coefficient is subject to a perturbation due to the delta deformation coefficient.

\section{$\delta$-deformed instanton solution} \label{sec:instdef}
\begin{figure}[h!]
	\centering
	\begin{subfigure}{0.45\textwidth}
	\includegraphics[width=0.95\linewidth]{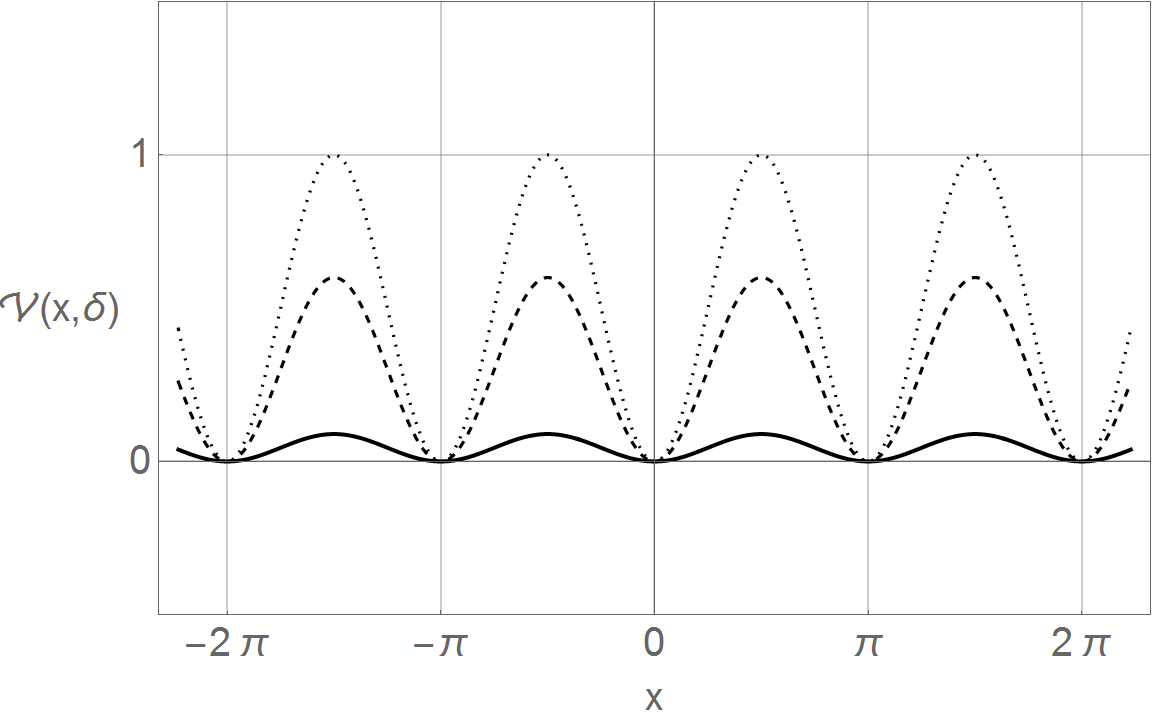}
	\caption{The deformed Mathieu potential in the Hermitian scenario, Eq.\@(\ref{eq:potencialz}). As expected, the higher the deformation parameter value, the lower the potential amplitude.}
	\label{fig:potencialdeformado}
    \end{subfigure}
\hfill
	\begin{subfigure}{0.45\textwidth}
	\includegraphics[width=0.95\linewidth]{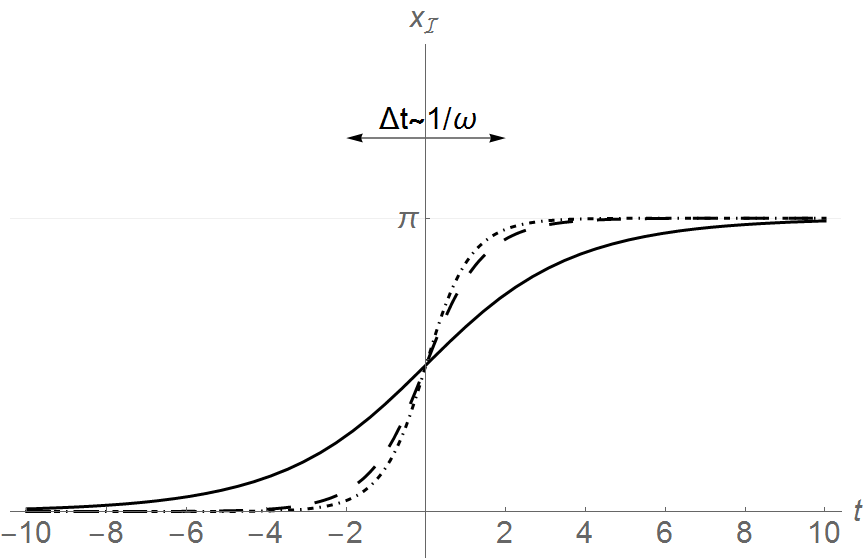}
	\caption{Real instantons in the Hermitian scenario, Eq.(\ref{eq:solution}). We have just observed that the height of the potential barrier decreases as we increase the value of $\delta$ (Fig.\@ \ref{fig:potencialdeformado}). This fact does make it smoother for the instanton to pass from one minimum to another. As $\delta \rightarrow 1$, the respective curve resembles a line whose slope is given by $\sqrt{2}(1-\delta^2)^{1/4}$. The instanton width $\Delta t$ varies with the value of $\delta$, being proportional to $1/\omega=2^{-1/2}(1-\delta)^{-1/4}$} 
	\label{fig:instanton3deltas}
    \end{subfigure}
    \caption{Effects of $\delta$ deformation parameter on (a) $\mathcal{V}(x,\delta)$ potential and (b) $x_\mathcal{I}(t)$ instanton solution. The different curves refer to different values of the non-Hermitian parameter: $\delta=0$ (dotted line), $\delta=0.8$ (dashed line) and $\delta=0.996$ (solid line). Wolfram Mathematica 11.3 was used to plot the graph.}
\end{figure}

We now turn to one-dimensional quantum mechanics, in which the partition function is a path integral,
\begin{equation}
\mathcal{Z}=\int \mathcal{D}x  \exp\left\{ -\int_{-\beta /2}^{\beta/2} dt \left( \frac{m}{2} \dot{x}^2 +\frac{\mathcal{V}(x, \delta)}{g^2} \right)\right\}.
\end{equation}
Setting $m=g=1$, the associated instanton equation $\dot{x}(t)=\sqrt{2\mathcal{V}(x(t), \delta)/(mg^2)}$ (see Refs.~\cite{bib:coleman,bib:livromarino,Tesenat}) has the solution
\begin{eqnarray} \label{eq:completesol}
  x(t) &=& -4 \mathrm{arctan}\left[\exp\left(-t \,\sqrt{2} (1-\delta^2)^{1/4}\right)\right.\nonumber\\
  &-& \left.\sqrt{
   1 + \exp\left(-t \, 2\sqrt{2}(1-\delta^2)^{1/4}\right)}\right]+4\pi C_1 ,\nonumber\\
   & & 
\end{eqnarray}
with $C_1 \in \mathbb{C}$. We define the real part of the solution as $\mathrm{Re }x(t)\equiv x_\mathcal{I}(t)$, which can be written as
\begin{equation} \label{eq:solution}
	x_\mathcal{I}(t)=2 \mathrm{arccot}\left(\exp\left(-t\sqrt{2\sqrt{1-\delta^2}}\right)\right),
\end{equation}
and its range is from $x_i=0$ to $x_f=\pi$, as shown in Fig. \ref{fig:instanton3deltas}. We are not interested in the imaginary part of the solution \ref{eq:completesol}, as it is constant. The instanton action $\left(S=\int_{x_i}^{x_f} dt \, m \dot{x}\right)$, see Refs.~\cite{bib:coleman,Tesenat}, is also influenced by the $\delta$ parameter,
\begin{equation} \label{eq:action}
S=\int_0^\pi \sqrt{2}(1-\delta^2)^{1/4}\sin(x)dx = 2\sqrt{2}\, (1-\delta^2)^{1/4}.
\end{equation}
Since the instanton action is related to singularities in the Borel plane, this amazing result denotes that one could manipulate the position of the singularities in the Borel plane by modifying the value of the non-Hermiticity factor, which characterizes the subject to be explored in a forthcoming paper.

Furthermore, in the region permeated by the instanton, there is a guarantee that the $\mathcal{PT}$ symmetry is not broken, because this is a closed scenario and there are still turning points~\cite{Bender:2021arxiv}. 

We've already discussed the reason for the restriction $\delta \ne 1$. But what would happen if this was allowed? The potential in (\ref{eq:deformed}) with $\delta=1$ is written $V(z,1)=\exp(iz)$. The solution of the instanton equation $\dot{z}=\sqrt{2 \, V(z,1)}$ exists, being written as $z=2i\ln(-i t \sqrt{2}/2)$, although it does not describe a typical instanton curve (Fig. \ref{fig:instanton3deltas}). First, we observe that it has a discontinuity in $t=0$. Furthermore, we can assert that putting $\delta=1$ does not produce any instanton because there are no saddle points in this case.

\begin{figure}[h!]
    \centering
    \includegraphics[width=0.95\linewidth]{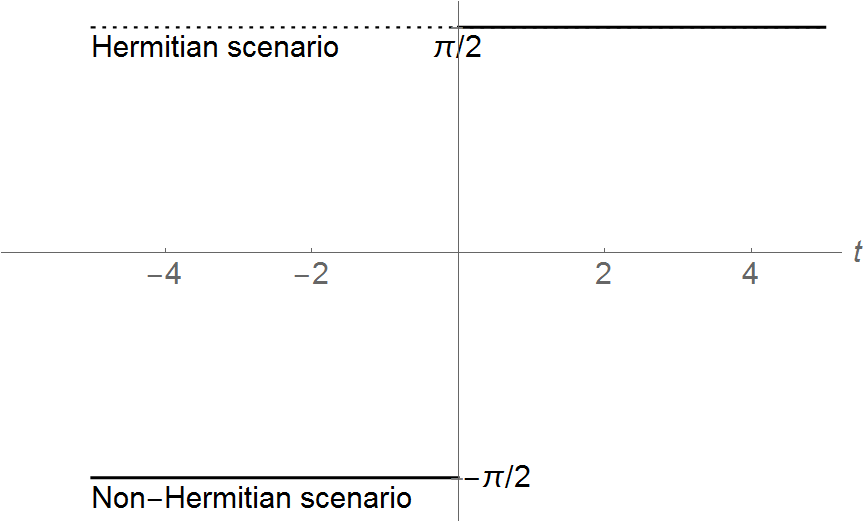}
    \caption{Solid line: Real part of the solution of $\dot{z}=\sqrt{2 \, V(z,1)}$ for the non-Hermitian scenario of the potential in Eq.\@ (\ref{eq:deformed}), with $\delta=1$. Dashed line: Real part of instanton solution for the Hermitian scenario, Eq.\@ (\ref{eq:solution}) with $\delta=1$, which appears as a constant line at $\pi/2$ for all $t$. Wolfram Mathematica 11.3 was used to plot the graph.}
    \label{fig:InstantonDelta1}
\end{figure}

We see that the two curves in Fig.\@ \ref{fig:InstantonDelta1} do not agree. This is expected, because the non-Hermitian scenario starts with the assumption $\delta=1$. On the other hand, the Hermitian scenario is obtained from the transformation (\ref{eq:vxialpha}) which is valid in the interval $\delta \in \left]-1,1\right[$ (see Eq.(\ref{alpha})). The limit $\delta\rightarrow 1$ is an extrapolation of the latter case and it does not need to be in agreement with the first scenario, as we observe in Fig.\@ \ref{fig:InstantonDelta1}. It does not bring the complete information, however. Then, one could naively think that the curves do not need to totally agree with each other, but at least they do in $t>0$.

For the non-Hermitian case, the only analytic solution for the instanton equation is for $\delta=1$, but this is not a true instanton. Recall that, in this scenario, there are no saddle points. Notwithstanding, we have shown that it is possible to investigate the effects of varying the $\delta \in \left]-1,1\right[$ deformation parameter --- we finally get true instanton solutions (see Fig.\@ \ref{fig:instanton3deltas}). This was done using the connection between non-Hermitian and Hermitian scenarios in section \ref{sec:connection}. In Appendix \ref{ap:A} we offer an alternative approach.

In the following sections, we will see how the behavior of several instantons is affected by the $\delta$ non-Hermiticity factor, not just one instanton, as we have done so far.

\section{Dilute gas of deformed instantons} \label{sec:dilute}
It is known that instantons are objects that appear in a short time interval~\cite{bib:Das}, having a width on order of $1/\omega$ (see Fig.\ref{fig:instanton3deltas}). It means that, for large $t=T$, the approximate solutions of the equation of motion are not just instantons and anti-instantons, but also strings of widely separated instantons and anti-instantons. In the case of periodic potentials, in which the Mathieu potential is understood, there are degenerate minima that allows us to observe the appearance of a number of instantons and anti-instantons freely moving on the real axis~\cite{bib:coleman,Zinn-Justin:1982aya}. Instantons can start from an initial position $x=j$ and go to the next one, $x=j+1$, while anti-instantons can go from $x=j$ to $x=j-1$. Thus, we write 
\begin{eqnarray}
    & &\braket{j_{+}|e^{-HT/\hbar}|j_{-}} =\left( \frac{\omega}{\pi \hbar}\right)^{1/2} \, \exp\left(-\frac{\omega T}{2}\right) \nonumber\\
    & & \times \sum_{n,\bar{n}=0}^\infty \frac{(K e^{-S/\hbar} T)^{n+\bar{n}}}{n! \;\bar{n}!} \delta_{n-\bar{n}+j_{+}+j_{-}}, \nonumber\\
    &  &
\end{eqnarray}
where $H$ is the Hamilonian, $\ket{j_{+}}$ and $\ket{j_{-}}$ are position eigenstates, $n$ is the number of instantons, $\bar{n}$ is the number of anti-instantons. If it were not for the small intervals containing instanton and anti-instanton, $\mathcal{V}''(x,\delta)$ would be equal to $\omega^2$ on the entire time axis. This characterizes the barrier penetration and that is where the factors before the sum come from. The correction of the formula that includes the influence of multi-instantons is within the summation, being attributed mainly to the $K$ factor. The expression $(T^{n+\bar{n}}/(n! \; \bar{n}!))$ is a consequence of the integration over the locations of the centers and $K$ is explicitly 
\begin{equation} \label{eq:K}
K=\left(\frac{S}{2\pi\hbar}\right)^{1/2} \left| \frac{\det(- \partial_t^2 + \omega^2)}{\det'(- \partial_t^2+\mathcal{V}''(x_\mathcal{I}))} \right|^{1/2},
\end{equation}
where the determinant is understood in four-dimensional Euclidean space, $x_\mathcal{I}$ is a single instanton, $S$ is the one-instanton action, $\omega$ is the one-instanton width and $\det'$ means that the zero eigenvalue is omitted when computing the determinant.

The terms of an exponential series $\sum_{k=0}^\infty y^k/ k!$ grow with $k$, for any fixed $y$, just until $k$ is on the order of $y$. Then, they begin to decrease fast. We can apply this to the sum in Eq.\@ (\ref{eq:K}) to conclude that the important terms are those for which the number of (anti-)instantons is less or approximately $K e^{-S/\hbar} \, T$. For small $\hbar$, it means that the density of (anti-)instantons is exponentially small when considering such important terms. Therefore, the average separation is huge. This is why this approach is called the dilute-gas approximation of instantons. The conditions on the validity of the average separation do not depend on $T$, as long as $T$ is sufficiently large.

From the instanton solution, Eq.\@ (\ref{eq:solution}), we identify $\omega$, see Fig.\@ \ref{fig:instanton3deltas}. The expression that describes the instanton action is in Eq.\@ (\ref{eq:action}). By comparison, we have $S=2\omega=2\sqrt{2}(1-\delta)^{1/4}$. Then, we can already note that the deformation parameter is strongly present even at the multi-instantons level.

The following section goes beyond the dilute instanton gas approximation and shows a case in which there is an interaction between a coupled instanton--anti-instanton pair.

\section{Application: Josephson junctions described by tilted $\delta$-deformed washboard potential} \label{sec:washboard}
The Josephson effect is a phenomenon that occurs when two or more superconductors are placed next to each other, with some potential barrier or restriction between them. This arrangement generates a supercurrent that flows through this device known as the Josephson junction. There are ways to build superconducting circuits based on the Josephson junction. They are candidates for developing quantum computing devices. As an example of this type, we have the  superconducting quantum interference device (SQUID)~\cite{bookWeiss}. 

In this context, the role of tunneling coordinate across the Josephson junction is played by the phase difference of the Cooper pair wave function. The standard model for describing this coupling is the resistively shunted junction (RSJ) model~\cite{bookWeiss},
\begin{eqnarray} \label{eq:RSJ}
C\left( \frac{\phi_0}{2\pi} \right)^2 \frac{d^2\psi}{dt^2} &+& \frac{1}{R}\left( \frac{\phi_0}{2\pi} \right)^2 \frac{d\psi}{dt} \nonumber \\ &+& E_J \sin\psi - I_{\textrm{ext}}\frac{\phi_0}{2\pi}=0, \nonumber \\ & & 
\end{eqnarray}
where $C$ and $R$ are respectively the capacitance and the effective shunt resistance of the junction, $\phi_0=h/2e$ is the flux quantum, $I_{\textrm{ext}}$ is the external bias current, and $E_J=I_c \phi_0/2\pi$ is the Josephson coupling energy, which is related to the $I_c$ maximum supercurrent supported by the junction.

There is a parallel between the equation of motion of this model, Eq. (\ref{eq:RSJ}), and the one corresponding to the motion of a Brownian particle of mass $M$ and position $X$ in the absence of fluctuations~\cite{bookWeiss}, given by
\begin{equation} \label{eq:browniana}
M\ddot{X}(t)+M\gamma \dot{X}(t)+\frac{\partial V(X)}{\partial X}=0, 
\end{equation}
where the potential is known as the tilted washboard, which is nothing more than the trigonometric-shape potential with a tilt. Note that we can insert the potential (\ref{eq:potencialz}) of the Hermitian scenario in this context if we include a term for the tilt and make $x=\pi X/X_0$. Thus, we write
\begin{eqnarray} \label{eq:washboard}
V(X)=-\frac{\sqrt{1-\delta^2}}{2}& & V_0\left[\cos\left(\frac{2\pi X}{X_0} \right)-\frac{1}{2}\right] \nonumber \\ & & - V_{\textrm{tilt}}\left(\frac{2\pi X}{X_0} \right),  
\end{eqnarray}
where and $\delta \in \left]-1,1\right[$, $V_0$, $V_{\textrm{tilt}} \in \mathbb{R}$. Comparing the equations (\ref{eq:RSJ}), (\ref{eq:browniana}) and (\ref{eq:washboard}), we see that there is an equivalence relation $(V_0/2)\sqrt{1-\delta^2} \; \hat{=} \; E_J$. A simple algebraic manipulation generates
\begin{equation}
|\delta| \; \hat{=}\; \sqrt{1-\frac{I^2_c h^2}{4V_0^2\pi^2 e^2}}.
\end{equation}
Therefore, the insertion of the tilted deformed Mathieu potential (\ref{eq:washboard}) in the Brownian motion equation (\ref{eq:browniana}) allows relating the non-Hermiticity factor $\delta$ with the $I_c$ Josephson supercurrent. 

\begin{figure}[h!]
    \centering
\begin{subfigure}{0.4\textwidth}
    \includegraphics[width=0.95\linewidth]{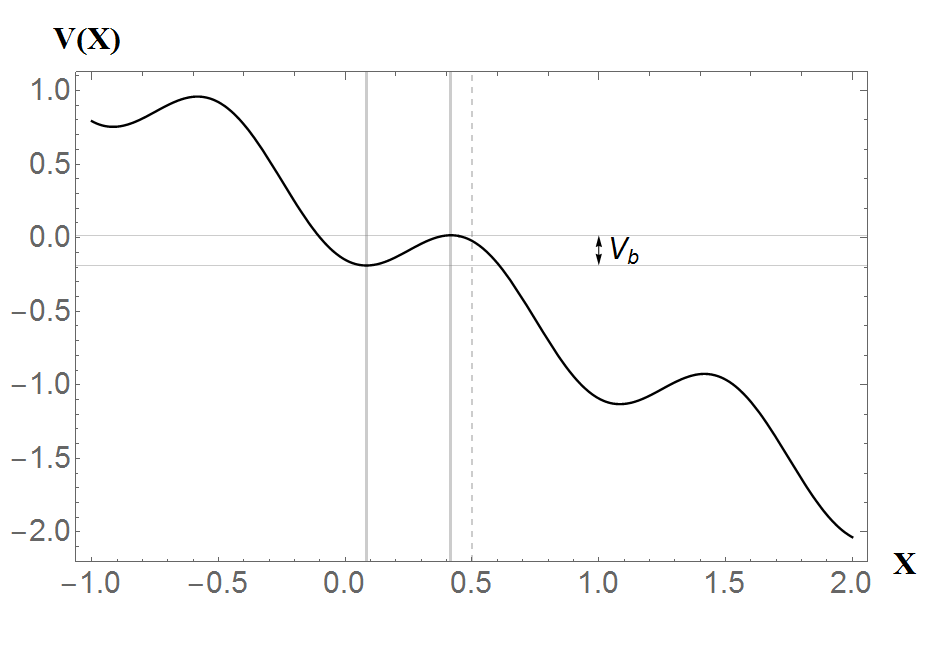}
    \caption{$\delta=0.8$, $X_0=1$, $V_{\textrm{tilt}}=0.15 J$, $V_0=1.0 J$. The critical points in the interval $0<X<X_0 /2$ are the minimum $X_1=(X_0/2\pi)\arcsin\left(V_\textrm{tilt}/(V_0\sqrt{1-\delta^2}/2)\right)$ and the maximum $X_2=(X_0/2)-X_1$. The height of the barrier is $V_b=V(X_2)-V(X_1)$, explicitly written in the Eq.\@ (\ref{eq:barrier}). Thick gridlines: minimum and maximum positions. Dashed gridline: $X_0/2$.}
    \label{fig:washboard}
\end{subfigure}
\hfill
	\begin{subfigure}{0.4\textwidth}
	\includegraphics[width=0.95\linewidth]{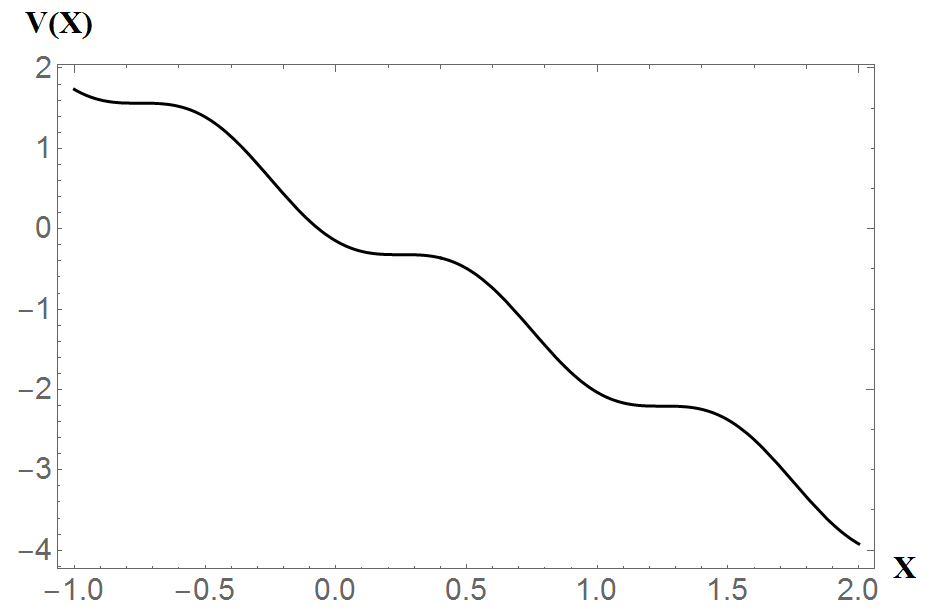}
	\caption{$\delta=0.8$, $V_{\textrm{tilt}}=0.3 J$, $V_0=1.0 J$, $X_0=1$. There are no cascade barriers.} 
	\label{fig:sembarreira}
    \end{subfigure}
\hfill
   \begin{subfigure}{0.4\textwidth}
	\includegraphics[width=0.95\linewidth]{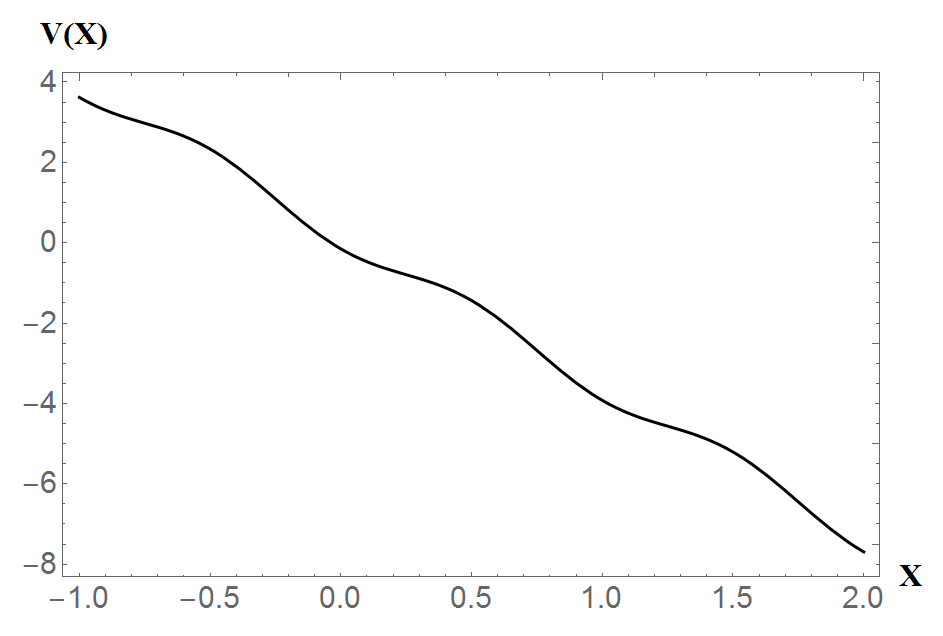}
	\caption{$\delta=0.8$, $V_{\textrm{tilt}}=0.6 J$, $V_0=1.0 J$, $X_0=1$. The slope is negative for all $X$.} 
	\label{fig:negativa}
    \end{subfigure}
\caption{The tilted $\delta$-deformed washboard potential (\ref{eq:washboard}) for (a) $|\delta|<\sqrt{1-4V_{\textrm{tilt}}^2/V_0^2}$, (b) $|\delta|=\sqrt{1-4V_{\textrm{tilt}}^2/V_0^2}$ and (c) $|\delta|>\sqrt{1-4V_{\textrm{tilt}}^2/V_0^2}$. Made with Wolfram Mathematica 11.3.}
\end{figure}

There are two distinct states in this current-biased Josephson junction: the zero-voltage state, which corresponds to the particle being trapped in a well, and the voltage state, that represents the particle sliding down the cascade of wells. Figure \ref{fig:washboard} makes it easy to see that the height of the barrier $V_b$ can be calculated by the difference between the value of the potential (\ref{eq:washboard}) at a maximum and its value at the previous minimum, such that
\begin{eqnarray} \label{eq:barrier}
V_b &=& 2\sqrt{\left(\frac{1-\delta^2}{4}\right)V_0^2-V_{\textrm{tilt}}^2} \nonumber \\ & & -2 V_{\textrm{tilt}} \arccos\left(\frac{V_{\textrm{tilt}}/V_0}{\sqrt{1-\delta^2}/2} \right).
\end{eqnarray}

From this result, we can see three different events. There are no cascade barriers for $|\delta|=\sqrt{1-4 V^2_{\textrm{tilt}}/V_0^2}$. The particle slides all the way down for $|\delta|>\sqrt{1-4 V^2_{\textrm{tilt}}/V_0^2}$. Finally, the proper condition for tunneling is $|\delta|<\sqrt{1-4 V^2_{\textrm{tilt}}/V_0^2}$. An interesting case within this last regime occurs when $|\delta|\ll \sqrt{1-4 V^2_{\textrm{tilt}}/V_0^2}$. In this limit, the bounce is described by a weakly bound instanton--anti-instanton pair. The distance between them is large compared to the width of an instanton. Thus, the action of the bounce is written as
\begin{equation}
S(T)/\hbar=2 S_{\mathcal{I}}/\hbar+W(T)-\epsilon T,
\end{equation}

where the action of a single instanton is
\begin{equation}
S_{\mathcal{I}} = \frac{M\gamma X_0^2}{2\pi\hbar}\left[ 1+\ln\left(\frac{\pi^2 V_0 \sqrt{1-\delta^2}}{M\gamma X_0^2 \omega_c} \right)\right].
\end{equation}
Also, the interaction term between the instantons at a distance $T$ is
\begin{equation}
W(T)= \frac{M\gamma X_0^2}{\pi\hbar}\ln[(\hbar\beta\omega_c /\pi)\sin(\pi T/\hbar\beta)],
\end{equation}
with $\omega_c$ taking the role of a reference frequency, and the last term $\epsilon T$ comes from the potential drop $\hbar \epsilon=2\pi  V_{\textrm{tilt}}$ on the distance $X_0$.

\section{Conclusion} 	\label{sec:conclusion}
In this work, we studied the implications of a $\mathcal{PT}$ symmetry deformation parameter $\delta$ in the transseries representation of the Mathieu partition function and in the instanton solution of the $\delta$-dependent Mathieu potential. First, we sought to establish a connection between the Hermitian and non-Hermitian scenarios. In order to do this, we worked with a non-Hermiticity factor restricted to the interval $\left]-1,1\right[$. We found that the deformation parameter is explicitly present in the non-perturbative saddle action and in the two formal series, perturbing the coupling coefficient. Therefore, the value of $\delta$ affects the transseries representation. Moreover, we showed that the width of the real instanton solution increases as the value of $\delta$  grows (tending to 1). Such a fact makes it smoother for the instanton to pass from one minimum to another. Our third observation was that the deformation parameter is present even at the multi-instanton scenario, characterizing a dilute gas of deformed instantons. We also studied the interesting case in which we make a parallel with the Josephson junction through a tilted deformed washboard potential. 
In the regime in which the bounce corresponds to a widely spaced instanton-anti-instanton pair, the $\mathcal{PT}$-deformation factor is present in the action of the bounce and affects the height of the potential barrier. 
We found that we can relate the non-Hermiticity factor with the Josephson supercurrent, indicating that the deformation might have a physical meaning. 
	
\backmatter
\bmhead{Acknowledgements}
N.\@ M.\@ Alvarenga thanks the Coordenação de Aperfeiçoamento de Pessoal de Nível Superior – Brasil (CAPES) – Finance Code 001 for financial support. 

J.\@ A.\@ Lourenço thanks the Fundação de Amparo à Pesquisa e Inovação do Espírito Santo -- Brasil (FAPES) -- PRONEM $N^o 503/2020$  for financial support.

We thank B.\@ W.\@ Mintz for suggesting the subject.


\begin{appendices}
\section{The role of the $\mathcal{PT}$ symmetry operator and the alternative approach using the original axis} \label{ap:A}
The first findings that certain $\mathcal{PT}$-symmetric, non-Hermitian Hamiltonians $H$ could have real spectra~\cite{bib:Bender1998,Bender:1999} were followed by a crucial investigation, the condition $[H,PT]=0$ alone is not enough to ensure the reality of the spectrum, because the associated Hilbert space has an indefinite inner product~\cite{bib:Brody2017}. So, what are the sufficient and necessary conditions to ensure a consistent $\mathcal{PT}$-symmetric quantum theory? 

The contribution provided by Refs.~\cite{bib:Mostafazadeh2002II,bib:Mostafazadeh2002III} started from a diagonalizable Hamiltonian with a discrete real spectrum to state that the reality of this spectrum is intrinsically linked to the existence of a positive-definite inner product ${\braket{ \cdot |\cdot }}_{+}$ that reproduces the self-adjoint Hamiltonian. This means that ${\braket{\psi|H\phi}}_{+}={\braket{H\psi|\phi}}_{+}$ for any pair of state vectors $(\psi,\phi)$.

The positive-definite inner product was determined in Ref.~\cite{bib:Mostafazadeh2002III} so that ${\braket{ \cdot |\cdot }}_{+}:=\braket{ \cdot |\eta_{+}\cdot }$ by a positive-defined operator $\eta_{+}$. This same operator also has a role in a similarity transformation, in which there is an invertible Hermitian $\rho=\sqrt{\eta_{+}}$ operator such that $H=\rho^{-1}h\rho$, with the Hamiltonian $H$ being $\mathcal{PT}$-symmetric non-Hermitian and $h$ Hamiltonian being Hermitian. In others words, this means that the reality of the spectrum of $H$ is also equivalent to the condition that $H$ is mapped to a Hermitian Hamiltonian $h$ through a similarity transformation. The so-called $\eta_{+}$ metric operator determines the physical Hilbert space and the observables of the quantum system. We see $H:\mathcal{H}\rightarrow\mathcal{H}$ as a linear operator acting in a Hilbert space $\mathcal{H}$ and $\eta_{+}:\mathcal{H}\rightarrow\mathcal{H}$ as a linear Hermitian automorphism. If the reader has wondered whether there is any investigation into non-diagonalizable pseudo-Hermitian Hamiltonians, the answer is affirmative and we quote the Refs.~\cite{bib:Mostafazadeh2002ndiag,bib:Mostafazadeh2004}.

An alternative to determine the conditions for the reality of the spectrum is in Ref.~\cite{bib:BenderComplex}, which proposes a $\mathcal{CPT}$-inner product that is equivalent to ${\braket{ \cdot |\cdot }}_{+}:=\braket{ \cdot |\eta_{+}\cdot }$ when choosing a particular $\eta_{+}$. The introduction of the symmetry $\mathcal{C}$, which is interpreted as a charge operator with eigenvalues $+1$ or $-1$ and commutes with the Hamiltonian $H$, causes $H$ to have a self-adjoint characteristic in the Hilbert space. On the other hand, if one decides to keep the inner product $\mathcal{PT}$ undefined and tries to formulate a quantum theory, the result is a non-linear quantum mechanics~\cite{bib:Brody2017}.

It is also important to discuss the extensibility of $\mathcal{PT}$ symmetry to the complex domain, because it is not immediate. Ref.~\cite{bib:Chakraborty} explains that the $SU(2)$ algebra can be deformed. Thus, their new generators are non-Hermitian, in the sense of the inner product in $\mathbb{C}^2$ and are constructed from a new set of bi-orthogonal vectors. In $SU(2)$, a spectrum generating algebra can be easily obtained through the Jordan-Schwinger map, with $J_m=\sum^2_{\mu,\nu=1} {(\sigma_m )}_{\mu \nu} a^\dagger_\mu a_\nu$, $m=1,2,3$, where $\sigma_m$ are the Pauli matrices and $a^\dagger_\mu ,  a_\nu$ belong to a set of boson operators. However, the Jordan-Schwinger realization of $SU_\gamma (2)$, which is the deformed one, does not directly produce a spectrum generating algebra with $J^\gamma_m$, $m=1,2,3$. It becomes necessary to introduce a new pair of ladder operators.~\footnote{This method can be used in a Hilbert space of dimension $n\geq 0$ with certain conditions~\cite{bib:Chakraborty}.}

The eigenfunctions of the operator $J_3^\gamma$ of $SU_\gamma(2)$ can be understood in the Fock space ($\mathcal{F}^2({\mathbb{C}}^2)$), which is a separable complex Hilbert space of integer functions~\cite{bib:Chakraborty}. The inner product that determines $\mathcal{F}^2({\mathbb{C}}^2)$ is
\begin{equation}
    \braket{\psi|\phi}=\int \int dW(\xi_1) dW(\xi_2) \; \psi(\xi_1,\xi_2) \overline{\phi(\xi_1,\xi_2)},
\end{equation}
where $dW(\xi_n)=\frac{1}{\pi} e^{-|\xi_n|^2} d(\textrm{Re}(\xi_n)) d(\textrm{Im}(\xi_n))$, $\xi_n$ being complex variables and $n=1,2$.
The $\mathcal{PT}$ symmetry is understood in a typical Fock space as a consequence of $\mathcal{C}_{(\vartheta,\eta,u)}$ (where $\vartheta,\eta,u$ are complex numbers) called weighted composition conjugation, 
\begin{equation}
    \mathcal{C}_{(\vartheta,0,1)} \psi(\xi)=\overline{\psi(\overline{-\xi})}.
\end{equation} 
Thus, we see that there is a solid ground concerning the role of the $\mathcal{PT}$ operator in terms of an underlying inner-product space.

In Sec.\@ \ref{sec:connection}, we have used the reparameterization $z=x+i\alpha(\delta)$ on the generalized $\mathcal{PT}$-symmetric Mathieu equation (\ref{eq:deformed}). It allowed us to connect Hermitian and non-Hermitian scenarios and to perform calculations in the referential given by this line parallel to the real axis, in the complex plane (see Fig.\ref{fig:reparametrizacao}). We choose this method by convenience, but this is not the only way to rewrite the $\mathcal{PT}$-symmetric Mathieu potential. We show in this Appendix that we can keep the complex variable $z$ unchanged in this process.


Consider multiplying the potential in Eq.(\ref{eq:deformed}) by a factor $\cos(i\epsilon)$,
\begin{equation}
\cos(i\epsilon) V(z,\delta)=\cos(i\epsilon)\cos(2z)+i\delta\cos(i\epsilon)\sin(2z).
\end{equation}
If we define $\delta$ such that $\delta\cos(i\epsilon)=-\sinh(\epsilon)$, or equivalently,
\begin{equation} \label{eq:deltaepsilon}
\delta=-\tanh(\epsilon). 
\end{equation}
This implies the same restriction on $\delta$ values that we have found in the section \ref{sec:connection}, that is, $\delta\in \; \left]-1,1\right[ \;$. We can write the deformed potential as
\begin{equation} \label{eq:ap}
\mathcal{V}(z,\delta)=\frac{\cos(2z+i\epsilon)}{\cos(i\epsilon)}.
\end{equation}
Thus, we can conclude that it is enough to restrict the values of the non-Hermiticity factor to the interval $\delta\in \; \left]-1,1\right[ \;$ to achieving the trigonometric form of the potential (Eq.\@ \ref{eq:ap}). To conveniently calculate the instanton solution of this potential, $\dot{z}(t)=\sqrt{2\mathcal{V}(z(t), \delta)}$, we can rewrite it as
\begin{equation}
\mathcal{V}(z,\delta)=\frac{2\sin^2(z+i\epsilon)}{\cos(i\epsilon)}.
\end{equation}
Thus, we find 
\begin{equation} \label{eq:apsolution}
z(t)=-i\epsilon+2\left\{ \mathrm{arctan}\left[\exp\left( 2t\sqrt{\mathrm{sech}(\epsilon)} \right) \right] \right\}.
\end{equation}

\begin{figure}[h!]
    \centering
    \includegraphics[width=0.95\linewidth]{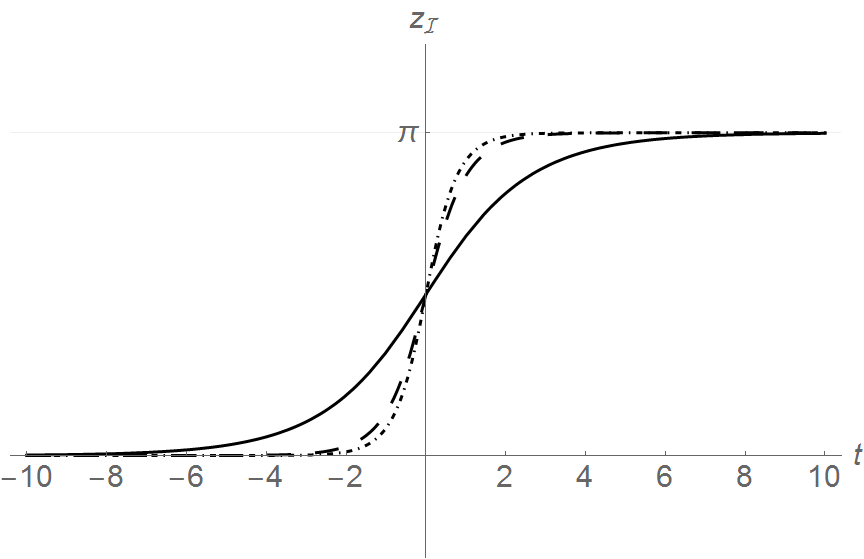}
    \caption{Real part of (\ref{eq:apsolution}) instanton solution, $z_{\mathcal{I}}\equiv\mathrm{Re}[z(t)]$, for $\epsilon=0$ (dotted line), $\epsilon=1.099$ (dashed line) and $\epsilon=3.106$ (solid line). Wolfram Mathematica 11.3 was used to plot the graph.}
    \label{fig:instantonap}
\end{figure}

We can find the $\epsilon$ values corresponding to the $\delta$ values used in Fig.\@ \ref{fig:instanton3deltas} through Eq.\@ \ref{eq:deltaepsilon}. The $\mathrm{Re} [z(t)]$ curves in Fig.\@ \ref{fig:instantonap} are identical to the $\mathrm{Re}[x(t)]$ curves plotted in Fig.\@ \ref{fig:instanton3deltas}).

\end{appendices}

\bibliography{refsfile}

\end{document}